\documentclass[usenatbib]{mn2e}
\pdfoutput=1
\usepackage[usenames,dvipsnames]{color}
\usepackage{graphicx}
\usepackage{subfig}
\usepackage{float}
\usepackage{natbib}
\usepackage{placeins}
\usepackage{caption}
\usepackage{times}

\newcommand{\mnras}{MNRAS}
\newcommand{\aap}{A\&A}

\newcommand{\apj}{ApJ}

\newcommand{\araa}{ARA\&A}
\newcommand{\apjs}{ApJS}

\title[XCS and the Planck Cluster Catalogue]
{The {\it XMM} Cluster Survey: Predicted overlap with the {\em Planck} Cluster Catalogue} 

\author[P.~T.~P.~Viana et al. (XCS collaboration)]{Pedro
  T.~P.~Viana,$^{1,2}$\thanks{E-mail: viana@astro.up.pt} Ant\'{o}nio da Silva,$^{1}$ 
  Elsa P.~R.~G.~Ramos,$^{1,2}$ Andrew R.~Liddle,$^{3}$ \cr
  E. J. Lloyd-Davies,$^{3}$ A.~Kathy Romer,$^{3}$ Scott~T.~Kay,$^{4}$ 
  Chris A.~Collins,$^{5}$ Matt Hilton,$^{6}$ \cr 
  Mark Hosmer,$^{3}$ Ben Hoyle,$^{7,8,9}$ Julian A. Mayers,$^{3}$ 
  Nicola Mehrtens,$^{3}$ \cr 
  Christopher J.~Miller,$^{10}$ Martin Sahl\'en,$^{11}$ 
  S.~Adam Stanford,$^{12,13}$ and John P.~Stott$^{5,14}$ \cr 
\\
\vspace*{-6pt} {\small \em $^{1}$Centro de Astrof\'{\i}sica, 
Universidade do Porto, Rua das Estrelas, 4150-762 Porto, Portugal}\\
\vspace*{-6pt} {\small \em $^{2}$Departamento de F\'{\i}sica e Astronomia, 
Faculdade de Ci\^{e}ncias, Universidade do Porto, Rua do Campo
Alegre, 687, 4169-007 Porto, Portugal}\\
\vspace*{-6pt} {\small \em $^{3}$Astronomy Centre, University of Sussex,
Falmer, Brighton BN1 9QH, UK}\\
\vspace*{-6pt} {\small \em $^{4}$Jodrell Bank Centre for Astrophysics, 
School of Physics and Astronomy, The University of Manchester, Manchester, M13 9PL, UK}\\
\vspace*{-6pt} {\small \em $^{5}$Astrophysics Research Institute,
Liverpool John Moores University, Twelve Quays House, Egerton Wharf,
Birkenhead CH41 1LD, UK}\\
\vspace*{-6pt} {\small \em $^{6}$ School of Physics \& Astronomy, 
University of Nottingham, NG7 2RD, UK}\\ 
\vspace*{-6pt} {\small \em $^{7}$Institut de Ci\'{e}ncies del Cosmos
  (ICCUB-IEEC), Departmento de F\'{\i}sica, Mart\'{\i} i Franqu\'{e}s 1,
  08034 Barcelona, Spain}\\ 
\vspace*{-6pt} {\small \em $^{8}$Institute of Cosmology and Gravitation, Dennis Sciama Building,
Burnaby Road, Portsmouth, PO1 3FX, UK}\\ 
\vspace*{-6pt} {\small \em $^{9}$Helsinki Institute of Physics, P.O. Box 64, FIN-00014 University of Helsinki, Finland}\\
\vspace*{-6pt} {\small \em $^{10}$Astronomy Department, University 
of Michigan, Ann Arbor, MI 48109, USA}\\
\vspace*{-6pt} {\small \em $^{11}$The Oskar Klein Centre for
  Cosmoparticle Physics, Department of Physics, Stockholm University, 
  SE-106 91 Stockholm, Sweden}\\ 
\vspace*{-6pt} {\small \em $^{12}$Physics Department, University of California, 
Davis, CA 95616, USA}\\ 
\vspace*{-6pt} {\small \em $^{13}$Institute of Geophysics and Planetary Physics, 
Lawrence Livermore National Laboratory, Livermore, CA 94551, USA}\\
\vspace*{-6pt} {\small \em $^{14}$Department of Physics, Institute for Computational Cosmology, Durham University, South Road, Durham, DH1 3LE, UK}\\
}

\begin{document}

\date{Accepted 2011 ???. Received 2011 ???; in original form 2011 September 8}

\pagerange{\pageref{firstpage}--\pageref{lastpage}} \pubyear{2012}

\maketitle

\begin{abstract}
We present a list of 15 clusters of galaxies, serendipitously detected by
the {\it XMM} Cluster Survey (XCS), that have a high probability of detection 
by the {\it Planck} satellite. Three of them already appear 
in the {\it Planck} Early Sunyaev--Zel'dovich (ESZ) catalogue. 
The estimation of the {\it Planck} detection probability assumes the flat
Lambda cold dark matter ($\Lambda$CDM) cosmology most compatible with 
7-year {\it Wilkinson Microwave Anisotropy Probe} ({\it WMAP7}) data. It takes
into account the XCS selection function and {\it Planck} sensitivity, as well 
as the covariance of the cluster X-ray luminosity, temperature, and
integrated comptonization parameter, as a function of cluster mass and
redshift, determined by the Millennium Gas Simulations. We also 
characterize the properties of the galaxy clusters in the final data release 
of the XCS that we expect {\it Planck} will have detected by the end of 
its extended mission. Finally, we briefly discuss possible joint applications 
of the XCS and {\it Planck} data. 
\end{abstract}

\section{Introduction}

The properties of the intergalactic medium within clusters of galaxies
can be studied using observational data obtained both in the X-ray and
sub-mm/mm wavebands. These methods are complementary and can be 
combined not only to better characterise the internal structure of galaxy
clusters, but also to derive the angular-diameter distance as a
function of redshift thus helping to constrain cosmological
parameters (see \citealt{CHR} for a review).

The study of galaxy clusters in X-rays, through the bremsstrahlung
emission and line emission of the intracluster medium (ICM), started
significantly earlier than in the sub-mm and mm wavelengths through
the Sunyaev--Zel'dovich (SZ) effect they produce. Large catalogues of
galaxy clusters assembled through their identification in X-rays
have been created since the {\em ROSAT} data became available
\citep{BCS,RDCS,BS,NORAS,BSHARC,NEP,ST,WARPS,Mullis,REFLEX,Burenin}
more than a decade ago, while similar catalogues assembled through
identification of galaxy clusters via the SZ effect are only now being 
generated \citep{Marriage,Vander,Williamson}. Among the X-ray and SZ
cluster surveys under way, the XCS: {\it XMM} Cluster Survey
\citep{XCS1,XCS2} and the {\it Planck} Cluster Survey \citep{Planck1} 
stand out for the very large number of galaxy clusters they are 
expected to detect. Both surveys are based on the identification 
of galaxy clusters in satellite data, from 
{\em XMM--Newton}\footnote{http://xmm.esac.esa.int/}
({\it XMM} hereafter) and the {\it Planck} satellite ({\it Planck}) respectively.

Although the SZ effect can be used on its own to constrain the
properties of the ICM, the usefulness of the information it contains
is maximized when X-ray data is also available \citep{CHR}. This is
particularly true in the case of the galaxy clusters expected to be
detected by {\it Planck}, because of the limited information content of the
expected SZ-related data, due to the low angular resolution and
sensitivity (for galaxy cluster studies) of {\it Planck}. Ideally, the
galaxy clusters for which {\it Planck} will be able to best characterise
their SZ signal would be re-observed with {\it XMM} or {\it Chandra}. 
However, it is not clear that this can be done on a large scale. It is 
therefore important to consider alternative sources of X-ray data for 
(at least some of) the clusters in the {\it Planck} Cluster Survey.

The aim of this paper is to identify and present the clusters of
galaxies in the XCS first data release, for which X-ray
temperatures have been estimated, that have the highest
probability of also being detected by {\it Planck}. We will also
characterise the distribution of the properties of the galaxy
clusters that we expect will be detected both by XCS and {\it Planck},
and discuss possible applications of the information contained in the
data gathered by both surveys on those galaxy clusters.

The structure of this paper is as follows. We start by reviewing the
characteristics of the XCS and the {\it Planck} Cluster Survey. Next,
we present the methodology we used to estimate the probability of
detection of an XCS galaxy cluster by {\it Planck}, and characterise
the expected overlap between the two surveys. We then discuss possible
applications of the data gathered by both XCS and {\it Planck} on the
galaxy clusters present in the joint sample.

\section{The {\it XMM} and {\it Planck} Cluster Surveys}
\label{XCSPlanck}

\subsection{{\it XMM} Cluster Survey}
\label{XCS}

The XCS collaboration is carrying out a systematic search for
serendipitous detections of clusters of galaxies in the outskirts of
publicly-available pointings in the {\it XMM} archive \citep{RVLM}. 
A fully automated procedure is in place, which allows the identification 
and classification of all sources in such pointings \citep{XCS1}. XCS 
cluster candidates are those sources that are classified as being more 
extended than the instrument point spread function. The available X-ray 
and optical data, including information from the literature, is then used to 
determine if such candidates can be confirmed as genuine galaxy 
clusters and to estimate their redshifts \citep{XCS2}. Finally, an X-ray 
spectroscopy analysis is carried out, using the archival {\it XMM} data, 
and the ICM temperature estimated if enough X-ray photon counts 
are available.

The XCS project is ongoing, but already $5,776$ {\it XMM}
pointings have been analysed, yielding a serendipitous cluster
candidate catalogue numbering $3,675$ entries \citep{XCS1}. 
Among these, we were able to optically confirm the presence of 
503 clusters of galaxies, of which 255 are new to the literature and 
356 are new X-ray discoveries, while 464 have a redshift estimate 
and 402 had their X-ray temperatures measured \citep{XCS2}. 
So far, the XCS covers a combined area close to $410$ square 
degrees suitable for cluster searching, taking into account 
overlapping and repeated exposures, and excluding regions of 
low Galactic latitude, the Magellanic Clouds, and pointings with 
very extended central targets \citep{XCS1}. Given that the XCS 
has only analysed the {\it XMM} observations performed up to 
mid-2009, and the mission lifetime has been extended until the 
end of $2014$, a conservative estimate for the final XCS area 
for cluster searching is about $600$ square degrees.

The XCS will be the largest catalogue of X-ray selected clusters ever compiled. 
It is expected to contain several thousand entries, more than 20 per cent of 
which will have an associated X-ray temperature derived from the
serendipitous data alone \citep{Sahlen}. The XCS will be accompanied
by a complete description of its selection function, making it a
valuable resource for the unbiased derivation of cluster scaling
relations and cosmological parameters. The XCS selection function is
derived by placing a sample of mock surface-brightness profiles into
real {\it XMM} Observation Data Files. These profiles were produced
assuming a simple symmetrical model for the ICM structure (for more 
details see \citealt{Sahlen}). This assumption has been validated by 
an investigation of the recovery rate of clusters with profiles 
drawn from the CLEF hydrodynamical simulation \citep{Kay07}, 
including clusters with cool cores and substructure, which has shown 
no significant change in the XCS detection efficiency \citep{XCS1}.

\subsection{{\it Planck} Cluster Survey}
\label{Planck}

The {\em Planck} satellite has nine frequency channels
ranging from 30 GHz to 857 GHz, with an angular resolution reaching 
5 arcmin for the frequency channels above 143 GHz. Its objective is to
characterise the temperature anisotropies and polarisation of the
cosmic microwave background radiation (CMBR). These include 
anisotropies produced via the SZ effect in the direction of galaxy 
clusters. Their characteristic frequency dependence and expected 
profiles can be used to extract cluster candidates and their 
integrated comptonization parameter. Most recent analyses suggest 
that during its original nominal 14-month mission, {\it Planck} will be 
able to detect around 2000 galaxy clusters \citep{Melin,Bart,Leach,Chamb}. 

The {\it Planck} Early Sunyaev--Zel'dovich (ESZ) catalogue has now been 
released, and contains 189 galaxy clusters detected with a signal-to-noise 
higher than six \citep{Planck2}, of which 20 were previously unknown. An 
extended mission for {\it Planck} is already underway, which is expected to 
roughly double the observing time with respect to the nominal {\it Planck} 
mission \citep{Planck1}. 

\section{Methodology}
\label{Method}

The probability of detection by {\it Planck} of an XCS galaxy cluster,
with X-ray luminosity $L$ and temperature $T$, is given by
\begin{equation}
P(>Y_{\rm min} | L, T)=\frac{P(>Y_{\rm min}, L, T)}{P(L, T)}\,,
\end{equation}
where we are assuming that all the clusters with an integrated
comptonization parameter, $Y$, above $Y_{\rm min}$, and only these,
are detected by {\it Planck}. This has been shown by \cite{Melin} to 
be a good approximation for the {\it Planck} sensitivity, as long 
as $Y_{\rm min}$ is allowed to vary as a function of the cluster angular
size. We use the information in fig. 2 of \cite{Melin} to model this 
dependence, as a function of the angular size of the cluster core radius, 
$\theta_c$, in the context of an isothermal $\beta$-model with $\beta=2/3$. 
Also from fig. 2 in \cite{Melin}, it can be inferred that the {\it Planck} 
detection limit for unresolved clusters, in terms of $Y$ defined 
within the virial radius, is around $1 \times 10^{-3}\;{\rm arcmin^{2}}$ 
for a signal-to-noise of 5 (nominal {\it Planck} mission), the minimum 
necessary for a reliable estimation of $Y$ from {\it Planck} data 
\citep{Melin}. 
However, an analysis of the characteristics of the {\it Planck} Early 
Sunyaev--Zel'dovich (ESZ) catalogue suggests that it is more appropriate 
to consider $1 \times 10^{-3}\;{\rm arcmin^{2}}$ as the {\it Planck} detection 
limit for unresolved clusters when $Y$ is defined within a sphere 
encompassing a density contrast 200 times the critical density, 
i.e. within $r_{200}$ \citep{Planck2}. Therefore, we assume 
\begin{equation}
\label{eq:ymin}
Y_{\rm min}=9 \times 10^{-4} \exp(0.7\,\theta_c^{0.8}) \;{\rm arcmin^{2}}
\end{equation}
to be valid for $\theta_c > 0.1$ in arcmin, with $Y_{\rm min}$ being 
the SZ integrated signal within $r_{200}$, and consider the {\it Planck} 
detection limit for unresolved clusters to be reached for lower values of 
$\theta_c$. This sensitivity also represents a reasonable extrapolation 
of the sensitivity achieved in the {\it Planck} ESZ survey \citep{Planck2}, 
when it is taken into account that the ESZ clusters that have signal-to-noise 
above six were detected in just 10 months of {\it Planck} data, the $Y$ 
values estimated by the {\it Planck} collaboration are within $5R_{500}$, 
and most of the ESZ clusters appear as extended in the {\em Planck} maps. 
With the extension of the {\it Planck} mission, the detection limit 
goes down proportionally by a factor of roughly $\sqrt{2}$, given that the 
extended mission has a duration that is approximately double that of the 
nominal mission. Thus, in the case of the extended mission, for each 
$\theta_c$, the associated $Y_{\rm min}$ is calculated by dividing the 
result one obtains using equation (\ref{eq:ymin}) by $\sqrt{2}$. The XCS 
selection function was estimated assuming the cluster structure 
to be well approximated by an isothermal $\beta$-model with $\beta=2/3$, 
plus a universal core radius of $160\,{\rm kpc}$ \citep{Sahlen}. Therefore, 
in order to be consistent, we make the latter assumption when calculating 
$Y_{\rm min}$, which means this quantity will be the same for all clusters 
at a given redshift.

\begin{table*}
\centering
\begin{tabular}{@{}llcrrrrrc@{}}
\hline
\multicolumn{1}{l}{Cluster ID} &
\multicolumn{1}{c}{Alternative name} & 
\multicolumn{1}{c}{$z$} & 
\multicolumn{1}{c}{$L^{d}_{\rm bol}$} &
\multicolumn{1}{c}{$L_{\rm bol}$} &
\multicolumn{1}{c}{$T^{d}_{X}$} & 
\multicolumn{1}{c}{$T_{X}$} & 
\multicolumn{1}{c}{$Y$} & 
\multicolumn{1}{c}{$P_{\rm full}$} \\ 
\hline
\vspace{0.1cm}
  XMMXCS J151618.6$+$000531.3 & MaxBCG J229.07472$+$00.08903 & $0.12$ & $4.4^{+0.1}_{-0.1}$ & $4.4^{+0.1}_{-0.1}$ & $5.4^{+0.1}_{-0.1}$ & $5.3^{+0.1}_{-0.1}$ & $0.3^{+0.1}_{-0.1}$ & $1.00/1.00$\\
\vspace{0.1cm}
  XMMXCS J104044.4$+$395710.4 & ABELL 1068 & $0.14$ & $8.4^{+0.2}_{-0.2}$ & $8.2^{+0.2}_{-0.2}$ & $3.5^{+0.1}_{-0.1}$ & $3.6^{+0.1}_{-0.1}$ & $0.4^{+0.1}_{-0.1}$ & $1.00/1.00$\\
\vspace{0.1cm}
  XMMXCS J030348.3$-$775241.3* & 1RXS J030344.4$-$775222 & $0.27$ & $16.2^{+0.4}_{-0.4}$ & $16.3^{+0.3}_{-0.4}$ & $8.7^{+0.3}_{-0.3}$ & $8.3^{+0.3}_{-0.3}$ & $1.0^{+0.2}_{-0.1}$ & $1.00/1.00$\\
\vspace{0.1cm}
  XMMXCS J122658.1$+$333250.9 & WARP J1226.9$+$3332 & $0.89$ & $47.9^{+1.2}_{-1.1}$ & $47.6^{+1.1}_{-1.1}$ & $11.1^{+0.5}_{-0.5}$ & $11.4^{+0.4}_{-0.4}$ & $1.8^{+0.3}_{-0.2}$ & $1.00/1.00$\\
\vspace{0.1cm}
  XMMXCS J133254.8$+$503153.1*& RBS 1283 & $0.28$ & $12.5^{+0.4}_{-3.7}$ & $12.7^{+0.3}_{-0.2}$ & $7.7^{+0.3}_{-0.4}$ & $7.3^{+0.4}_{-0.3}$ & $0.8^{+0.1}_{-0.1}$ & $1.00/1.00$\\
\vspace{0.1cm}
  XMMXCS J111515.6$+$531949.5 & SDSS J1115$+$5319 CLUSTER & $0.47$ & $20.5^{+0.1}_{-0.1}$ & $20.5^{+0.1}_{-0.1}$ & $5.4^{+1.5}_{-0.9}$ & $8.3^{+0.4}_{-0.4}$ & $1.1^{+0.2}_{-0.2}$ & $0.99/1.00$\\
\vspace{0.1cm}
  XMMXCS J090101.5$+$600606.2 & MaxBCG J135.25325$+$60.10133 & $0.29$ & $19.1^{+3.9}_{-3.2}$ & $16.7^{+3.1}_{-2.8}$ & $5.9^{+2.9}_{-1.4}$ & $7.7^{+0.6}_{-0.6}$ & $1.0^{+0.2}_{-0.2}$ & $1.00/1.00$\\
\vspace{0.1cm}
  XMMXCS J113020.3$-$143629.7* & ABELL 1285 & $0.11$ & $5.7^{+4.5}_{-1.7}$ & $4.5^{+1.1}_{-1.0}$ & $5.4^{+0.7}_{-0.7}$ & $5.0^{+0.5}_{-0.4}$ & $0.3^{+0.1}_{-0.1}$ & $1.00/1.00$\\  
\vspace{0.1cm}
  XMMXCS J033049.7$-$522836.5 & ABELL 3128 NE & $0.44$ & $20.9^{+0.2}_{-0.2}$ & $20.9^{+0.1}_{-0.1}$ & $4.5^{+0.1}_{-0.1}$ & $4.9^{+0.1}_{-0.1}$ & $0.8^{+0.1}_{-0.1}$ & $0.71/1.00$\\
\vspace{0.1cm}
  XMMXCS J021440.9$-$043321.9 & ABELL 0329 & $0.14$ & $2.8^{+3.4}_{-1.6}$ & $3.1^{+0.6}_{-0.5}$ & $4.5^{+0.1}_{-0.1}$ & $4.5^{+0.1}_{-0.2}$ & $0.2^{+0.1}_{-0.1}$ & $0.76/1.00$\\	 	
\vspace{0.1cm}
  XMMXCS J004624.5$+$420429.5 & RX J0046.4$+$4204 & $0.30$ & $7.0^{+0.3}_{-0.3}$ & $7.0^{+0.3}_{-0.3}$ & $6.9^{+0.6}_{-0.6}$ & $6.0^{+0.3}_{-0.3}$ & $0.5^{+0.1}_{-0.1}$ & $0.19/0.98$\\
\vspace{0.1cm}
  XMMXCS J141832.3$+$251104.9 & WARP J1418.5$+$2511 & $0.29$ & $6.3^{+0.5}_{-0.5}$ & $6.4^{+0.4}_{-0.5}$ & $6.4^{+0.4}_{-0.4}$ & $5.9^{+0.3}_{-0.3}$ & $0.4^{+0.1}_{-0.1}$ & $0.13/0.94$\\
\vspace{0.1cm}
  XMMXCS J123019.6$+$161634.1 & NSC J123020$+$161652 & $0.20$ & $4.6^{+0.8}_{-0.7}$ & $4.0^{+0.7}_{-0.5}$ & $4.3^{+0.6}_{-0.5}$ & $4.7^{+0.4}_{-0.3}$ & $0.3^{+0.1}_{-0.1}$ & $0.25/$0.90\\
\vspace{0.1cm}
  XMMXCS J121744.6$+$472921.5 & 400d J1217$+$4729 & $0.27$ & $23.2^{+13.2}_{-10.9}$ & $8.4^{+6.1}_{-3.4}$ & $9.8^{+6.6}_{-3.7}$ & $6.1^{+1.3}_{-1.0}$ & $0.5^{+0.4}_{-0.2}$ & $0.64/$0.85\\
  XMMXCS J095343.6$+$694735.0 & 400d J0953$+$6947 & $0.21$ & $1.0^{+3.0}_{-0.7}$ & $3.6^{+1.7}_{-1.3}$ & $5.7^{+1.1}_{-0.7}$ & $4.7^{+0.6}_{-0.5}$ & $0.3^{+0.1}_{-0.1}$ & $0.22/$0.56\\
\hline
\end{tabular}
\caption{All clusters in the XCS250 sample and $P_{\rm full}(>Y_{\rm min} | L, T) > 0.5$ in the case of the {\it Planck} 
extended mission. The bolometric ([0.05, 100] keV band) luminosities, $L^{d}_{\rm bol}$ and $L_{\rm bol}$, X-ray 
temperatures, $T^{d}_{X}$ and $T_{X}$, and integrated  comptonization parameter, $Y$, are defined 
within $r_{200}$ and have units of $10^{44}\;{\rm erg\,s^{-1}}$, ${\rm keV}$ and $10^{-4}\;h^{-2}\;{\rm Mpc^{2}}$,  
respectively. The uncertainty in the estimation of these quantities has a probability distribution which is close to 
log-normal, and the interval of variation presented corresponds to the 68\% confidence interval. The quantities 
($L^{d}_{\rm bol}$, $T^{d}_{X}$) and ($L_{\rm bol}$, $T_{X}$) differ in that the estimation of the latter takes into 
account the (prior) assumptions made in this work, with respect to the cosmological model, cluster scaling 
relations and XCS selection function. The probability of {\it Planck} detection, $P_{\rm full}(>Y_{\rm min} | L, T)$, 
is shown in the last column for both the nominal and extended missions (nominal/extended). The clusters with an 
asterisk next to their XCS name already appear in the {\it Planck} Early Sunyaev--Zel'dovich (ESZ) catalogue 
\citep{Planck2}. More information about the galaxy clusters in this table can be found in Mehrtens et~al. (2011). 
Several of these clusters have more than one alternative name. For example, MaxBCG J229.07472$+$00.08903 
is also known as Abell 2050 and RXC J1516.3$+$0005. Therefore, in order to be succinct, for each cluster in this 
table, we chose to present only the alternative name associated with the galaxy cluster that appears closest to the 
location of the XCS cluster when a query is submitted to NED (http://ned.ipac.caltech.edu/).
} 
\label{tab:xcsplancktable}
\end{table*}

In order to determine $P(>Y_{\rm min} | L, T)$, we need to know
both $P(>Y_{\rm min}, L, T)$ and $P(L, T)$ as a function of
redshift. These quantities can be obtained by integrating the joint 
probability function $P(L, T, Y)$, as follows
\begin{eqnarray}
P(>Y_{\rm min}, L, T) = \int_{Y_{\rm min}}^{\infty}P(L, T, Y){\rm d}Y\,,
\end{eqnarray}
and 
\begin{eqnarray}
P(L, T) = \int_{0}^{\infty}P(L, T, Y){\rm d}Y\,.
\end{eqnarray}

The full characterization of $P(L, T, Y | M)$, as a function of redshift, 
has been done only by \cite{Stanek}, using the data generated by the 
Millennium Gas Simulations (MGS). Therefore, we use their 
results\footnote{Note that the units of $Y$ should be taken to be 
$h^{-1}\;{\rm Mpc^{2}}$ in the derived MGS scaling relations between 
this quantity and cluster mass (G. Evrard and R. Stanek, private 
communication).}, namely those derived from the MGS with 
pre-heating, to estimate $P(L, T, Y)$ through
\begin{eqnarray}
P(L, T, Y) = \int_{M_{\rm min}}^{M_{\rm max}}P(L, T, Y | M)P(M){\rm d}M\,,
\end{eqnarray}
where
\begin{equation}
P(M) = \frac{n(M)}{\int n(M){\rm d}M}\,,
\end{equation}
with the integration going from 
$M_{\rm min}=5 \times 10^{13}\,h^{-1}\;{\rm M}_\odot$ to 
$M_{\rm max}=5 \times 10^{15}\,h^{-1}\;{\rm M}_\odot$. The lower
limit coincides with the MGS mass cut-off and $h$ is the value of the
Hubble constant in units of $100\;{\rm km\,s^{-1}\,Mpc^{-1}}$. The 
halo mass $M$ is defined as that which is contained within a sphere 
encompassing a density contrast 200 times the critical density, i.e. 
within $r_{200}$. All other cluster properties mentioned in this work, namely 
those that appear in Table~1 and Fig.~3, also refer to $r_{200}$.
The mass function, $n(M)$, is derived following \cite{jetal} and \citet{HK}, 
for a spatially-flat Cold Dark Matter cosmology with a spectrum 
of primordial adiabatic Gaussian density perturbations and the 
presence of a cosmological constant, $\Lambda$. We assume 
$\Omega_{\rm c} = 0.23$, $\Omega_{\rm b} = 0.04$, 
$\Omega_{\Lambda}=0.73$, $n_{\rm s} = 0.97$, $\sigma_8 = 0.81$ 
and $h= 0.70$, motivated by the WMAP-7 results \citep{K7}.

Both the X-ray luminosity and temperature of the XCS clusters are
measured with some associated uncertainty. What we then need to 
know is the probability of an XCS cluster having some X-ray luminosity,
$L$, and temperature, $T$, given the measured values, respectively
$L_{\rm obs}$ and $T_{\rm obs}$, for those quantities, i.e. $P[(L, T)
  | (L_{\rm obs}, T_{\rm obs})]$. This probability is different for each
XCS cluster, and its effect on the probability of detection by 
{\it Planck} of an XCS galaxy cluster, with measured $L_{\rm obs}$ 
and $T_{\rm obs}$, can be taken into account by marginalizing over 
$L$ and $T$, as follows
\begin{eqnarray}
\nonumber P_{\rm full}(>Y_{\rm min} | L_{\rm obs}, T_{\rm obs}) =
\int_{0}^{\infty}\int_{0}^{\infty}P(>Y_{\rm min} | L, T)\times\\  
P[(L, T) | (L_{\rm obs}, T_{\rm obs})]{\rm d}T{\rm d}L\,.
\end{eqnarray}

Naively, for each XCS galaxy cluster, the quantity 
$P[(L, T) | (L_{\rm obs}, T_{\rm obs})]$ would simply be taken to be 
the joint probability distribution of the estimated $L_{\rm obs}$ 
and $T_{\rm obs}$ values, based on the data collected for that cluster. 
This would be the case if we assumed we had no (prior) knowledge 
about the properties of the underlying galaxy cluster population and of 
the way we built the cluster sample under analysis, i.e. of the sample 
selection function. However, we not only already had to make 
assumptions about the underlying galaxy cluster population, in order 
to be able to estimate $P(>Y_{\rm min} | L, T)$, but we also believe we 
know how the XCS selection function, $f_{\rm XCS}(L, T)$, behaves. 
Therefore, we need to include these prior assumptions in the estimation 
of $P[(L, T) | (L_{\rm obs}, T_{\rm obs})]$. This can be done, for each 
XCS galaxy cluster, by multiplying the joint probability distribution of the 
estimated $L_{\rm obs}$ and $T_{\rm obs}$ values, based on the data 
collected for that cluster, by the probability of a cluster with $L$ and $T$ 
being present in the XCS catalogue, which is equal to the product of 
$f_{\rm XCS}(L, T)$ and $P(L, T)$, and by a re-normalization constant 
that ensures the integral of $P[(L, T) | (L_{\rm obs}, T_{\rm obs})]$ 
over all possible $L$ and $T$ is unity. 

In the following section, we will use the methodology just described 
to determine the probability of Planck detection for galaxy clusters in 
the XCS first data release, and to characterize the expected overlap 
between the XCS and {\it Planck} Cluster Survey. This methodology 
can, in principle, be applied to any two or more cluster surveys, for 
example eROSITA \citep{eROSITA}, DES \citep{DES} or 
SPT \citep{SPT}, in order to derive the expected number and  
properties of the objects common to those surveys being considered.

\section{Results and discussion}
\label{Results and discussion}

We have calculated $P_{\rm full}(>Y_{\rm min} | L, T)$ only for those 
galaxy clusters in the XCS first data release which have a minimum of 
250 X-ray photon counts, as otherwise the estimates of the X-ray 
observables would be too uncertain. Further, we have only considered 
clusters which have a X-ray temperature in excess of 2 keV and a 
redshift in the interval $0.1 < z < 1.0$ (the XCS selection function has 
not yet been evaluated outside this interval). Hereafter, these clusters 
will be referred to as constituting the XCS250 sample. In Table~1, all 
XCS250 clusters which have $P_{\rm full}(>Y_{\rm min} | L, T) > 0.5$ are 
presented. Note that even if this probability threshold was decreased to 
0.01, only 7 (13 for the extended mission) more clusters would be 
added to Table~1. 

Clearly, {\it Planck} will be able to detect only the closest, most 
luminous and hot clusters in the XCS. This can be also seen in 
Fig.~1, where we plot the minimum X-ray temperature, marginalized 
over X-ray luminosity and as a function of redshift, that an XCS cluster 
with over 250 X-ray photon counts needs to have in order to be detected 
by {\it Planck} with 0.5 probability, for both the nominal and extended 
missions. Its substantial increase with redshift is a consequence of the 
characteristics of the {\it Planck} Cluster Survey selection function, 
given that the minimum cluster mass, and hence X-ray temperature and 
luminosity, above which {\it Planck} can detect galaxy clusters grows 
strongly with redshift \citep{Melin,Planck2}. It was therefore not a surprise 
to find that all the clusters in Table~1 had already been detected in 
previous surveys \citep{XCS2}. Interestingly, the clusters
\mbox{XMMXCS J030348.3$-$775241.3 (ie. PLCKESZ G294.66$-$37.02)}, 
\mbox{XMMXCS J133254.8$+$503153.1 (ie. PLCKESZ G107.11$+$65.31)}, 
\mbox{XMMXCS J113020.3$-$143629.7 (ie. PLCKESZ G275.21$+$43.92)},
are already present in the {\it Planck} Early Sunyaev--Zel'dovich (ESZ) 
catalogue \citep{Planck2}, and have been found to have values for 
$Y$ \citep{Planck1}, after re-scaling to the same cluster radius, that 
are compatible within the uncertainties with those presented in Table~1. 
As expected, none of the XCS250 clusters not in Table~1 appear in the 
ESZ catalogue.  

The methodology developed in the previous section can also be used 
to estimate the probability distributions for the bolometric luminosity, 
$L_{\rm bol}$, X-ray temperature, $T_{X}$, and integrated comptonization 
parameter, $Y$, associated with any XCS cluster. We have found that such 
distributions are close to log-normal, and calculated for each cluster in 
Table~1 the most probable values for $L_{\rm bol}$, $T_{X}$ and $Y$, 
as well as the respective 68\% confidence limits. In Table~1, 
the quantities $L^{d}_{\rm bol}$ and $T^{d}_{X}$ are also shown. They 
differ from $L_{\rm bol}$ and $T_{X}$ in that they were derived solely 
based on an analysis of the XCS data \citep{XCS2}. Given that the estimation 
of $L_{\rm bol}$ and $T_{X}$ also depends on the (prior) assumptions regarding 
the cosmological model, cluster scaling relations and XCS selection function, 
the differences between the two sets of quantities gives an indication of the 
tension between the cluster data and the theoretical framework considered, 
which we discuss next.

The assumed cosmological model predicts steeply decreasing 
cluster mass, temperature and luminosity functions, and thus this prior 
tends to make $L_{\rm bol}$ and $T_{X}$ smaller than the equivalent 
data-based estimates. For example, this effect contributed to the significant  
reductions in the expected values for the bolometric luminosity and X-ray 
temperature of cluster XMMXCS J121744.6$+$472921.5, given that a cluster 
with a bolometric luminosity of $2.3\times10^{45}\,{\rm erg}\,{\rm s}^{-1}$ 
and X-ray temperature of 9.8 keV would be so massive, and thus so rare, 
that it would be highly unlikely to be detected in the present cosmological 
volume probed by the XCS. However, how small $L_{\rm bol}$ and $T_{X}$ 
can be, relative to their equivalent data-based estimates, will depend strongly 
on the magnitude of the uncertainties in the latter. For example, although 
the cluster XMMXCS J122658.1$+$333250.9 has higher $L^{d}_{\rm bol}$ 
and $T^{d}_{X}$ than XMMXCS J121744.6$+$472921.5, the low uncertainty 
in its X-ray temperature data-based estimate does not allow the prior on the 
cosmological model to lead to a prediction for the ratio $T_{X}$/$T^{d}_{X}$ 
as low as in the case of XMMXCS J121744.6$+$472921.5.

The XCS selection function has an opposite, 
but much smaller, effect to that produced by the assumed cosmological 
model: given that the lower the bolometric luminosity of a cluster the more 
unlikely is its detection by the XCS, the effect of the XCS selection function 
is to predict $L_{\rm bol}$ to be (slightly) higher than $L^{d}_{\rm bol}$. 
Note that the dependence of the XCS selection function on the cluster 
X-ray temperature is not monotonic \citep{Sahlen}, and thus it is not 
possible to predict beforehand if just taking into consideration the XCS 
selection function will result in a value for $T_{X}$ that is higher or lower 
than $T^{d}_{X}$. 

Finally, the cluster scaling relations, here assumed to be those derived 
from the pre-heating MGS, can also have a very strong effect on the 
estimates for $L_{\rm bol}$ and $T_{X}$, by forcing these quantities to 
comply with them. This may result in a higher $T_{X}$ with respect to 
$T^{d}_{X}$, as happened to clusters XMMXCS J111515.6$+$531949.5 
and XMMXCS J090101.5$+$600606.2, or in a lower $T_{X}$ with respect 
to $T^{d}_{X}$, as was the result for the cluster 
XMMXCS J095343.6$+$694735.0. Essentially, the former have a value 
for $L^{d}_{\rm bol}$ that is significantly higher than the scaling 
relations predict given their $T^{d}_{X}$, while for the latter the situation 
is reversed.
 
\begin{figure}
\centering
\includegraphics[width=\linewidth]{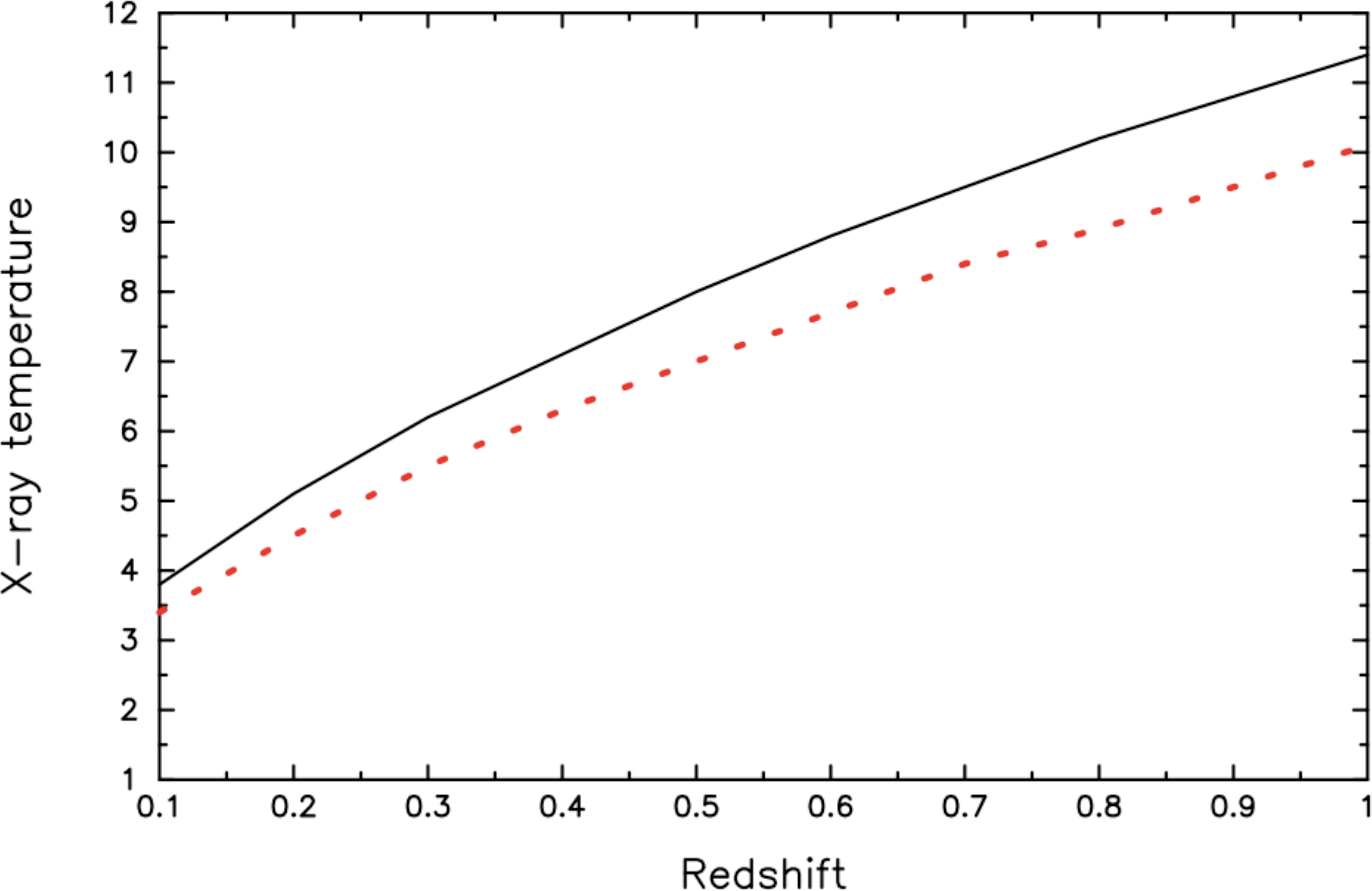}
\caption{Minimum X-ray temperature, as a function of redshift, that an XCS 
cluster with over 250 X-ray photon counts needs to have in order to be 
detected by {\it Planck} with 0.5 probability (nominal 
mission - black/full, extended mission - red/dashed).}
\label{fig:mintemp}
\end{figure}

Assuming that the final XCS catalogue will correspond to effectively
searching a sky area of $600$ square degrees for serendipitous
clusters, we have generated 20,000 random mock catalogues of the
expected overlap between the XCS and {\it Planck} Cluster Survey,
under the assumptions previously described. In Fig.~2 we plot the
distribution thus obtained for the expected total number of galaxy
clusters, with X-ray temperatures in excess of 2 keV and in the
redshift interval $0.1 < z < 1.0$, as well as enough X-ray photon
counts ($>$ 250) for their X-ray temperature to be estimated from the
serendipitous X-ray data alone, that will be simultaneously detected 
by the XCS and {\it Planck} Cluster Survey. Unfortunately, the 
expected overlap has a most probable size of only 7 
(3 to 13 with 95 per cent confidence) clusters for the nominal 
{\it Planck} mission, although it increases to 15 
(8 to 23 with 95 per cent confidence) clusters for the 
extended {\it Planck} mission. However, this calculation is 
sensitive to the assumptions made regarding the fiducial 
cosmology, the normalization and covariance of the cluster 
scaling relations, and the characteristics of the XCS 
selection function and {\it Planck} sensitivity. For example, 
increasing/decreasing the assumed value for 
$\sigma_{8}$ by 5 per cent induces an increase/decrease of the 
expected number of overlapping clusters to 12/4 (23/9 for the 
extended mission). In Fig.~3, we plot the mean distribution of the 
properties of the galaxy clusters that appear in the 20,000 random 
mock catalogues of the expected overlap between the XCS and 
{\it Planck} Cluster Survey, both for the nominal and extended 
missions, as well as the properties of the galaxy clusters presented 
in Table~1.

\begin{figure}
\centering
\includegraphics[width=\linewidth]{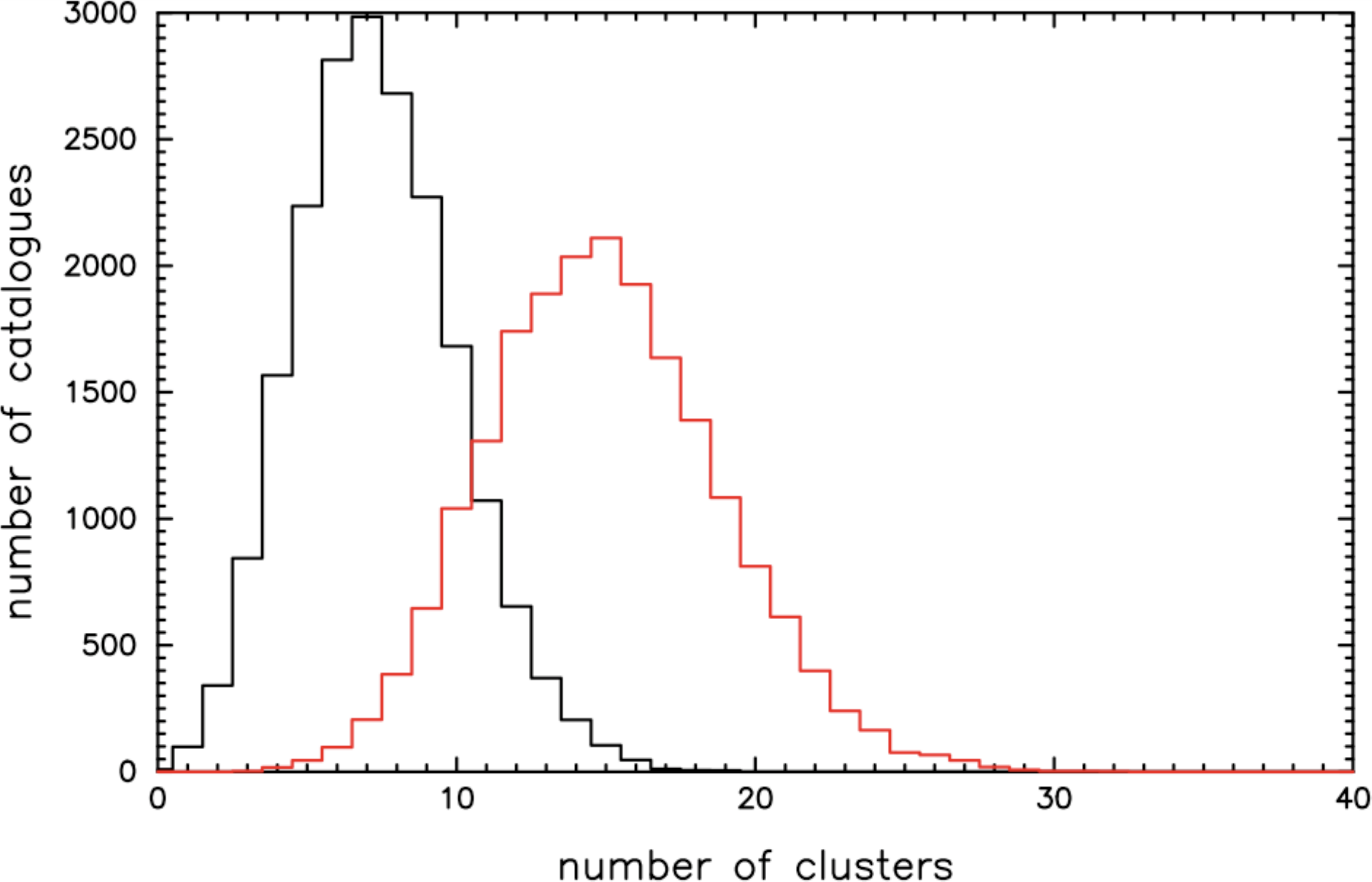}
\caption{Probability distribution of the expected total number 
of galaxy clusters, with X-ray temperatures in excess of 2 keV 
and in the redshift interval $0.1 < z < 1.0$, that will be 
simultaneously detected by the XCS (with more than 250 X-ray 
photon counts) and {\it Planck} Cluster Survey over 600 sq. deg.
of sky (nominal mission - black, on the left, extended 
mission - red, on the right).}
\label{fig:numcluster}
\end{figure}

\begin{figure*}
\centering
\subfloat[]{\label{fig:zclusters}
\includegraphics[width=8cm]{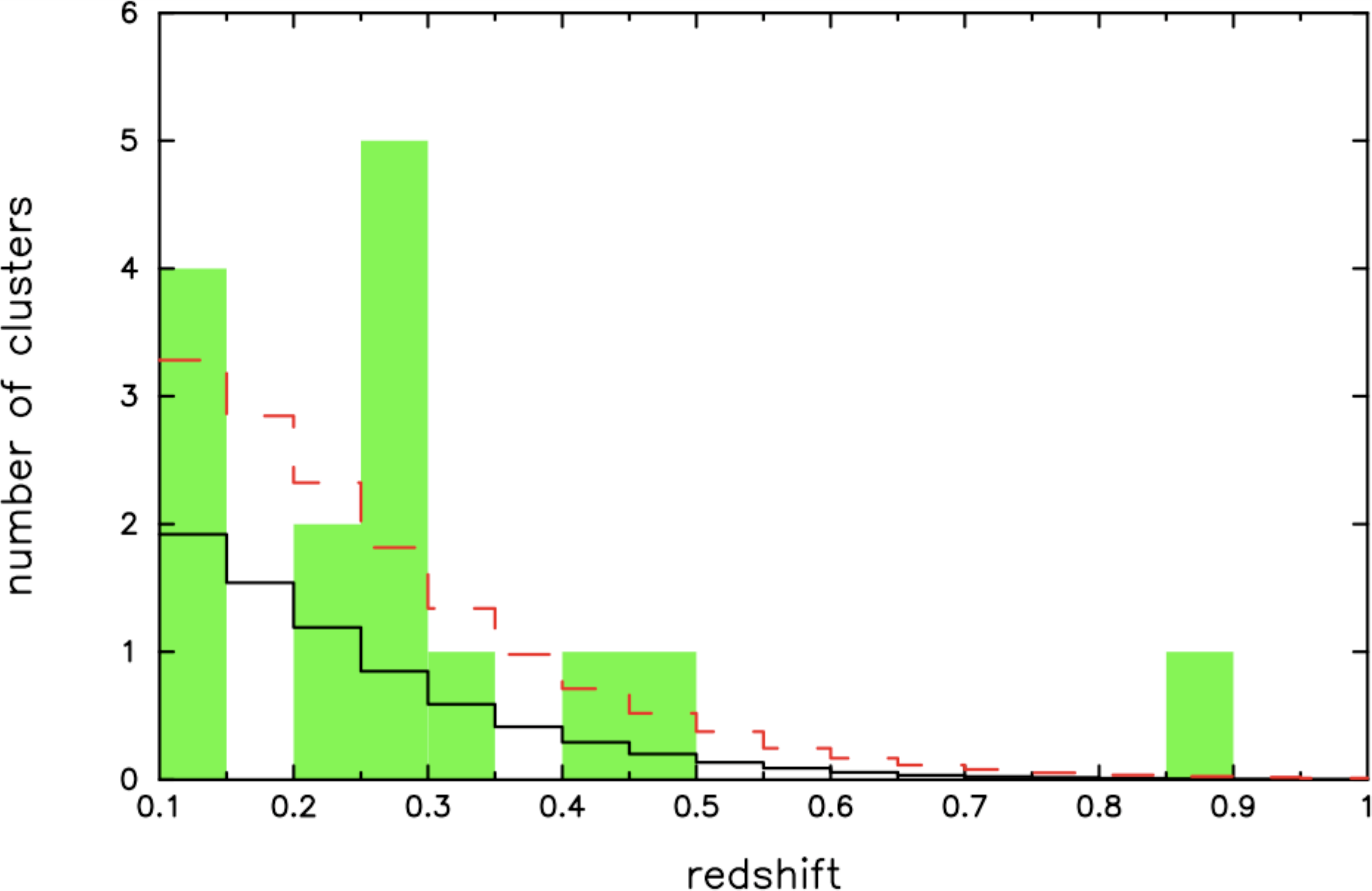}} 
\subfloat[]{\label{fig:Tclusters}
\includegraphics[width=8cm]{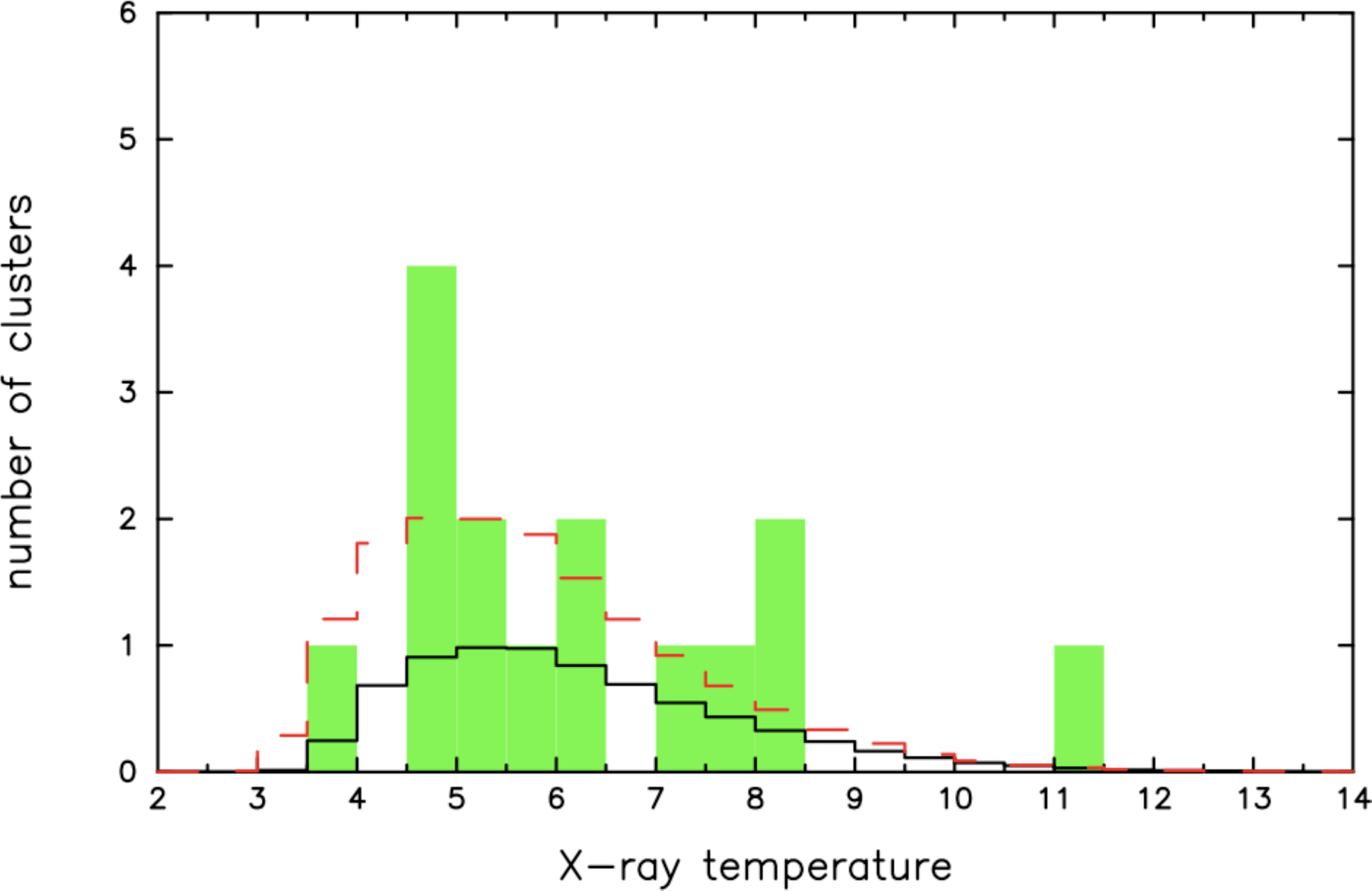}} \\
\subfloat[]{\label{fig:Lclusters}
\includegraphics[width=8cm]{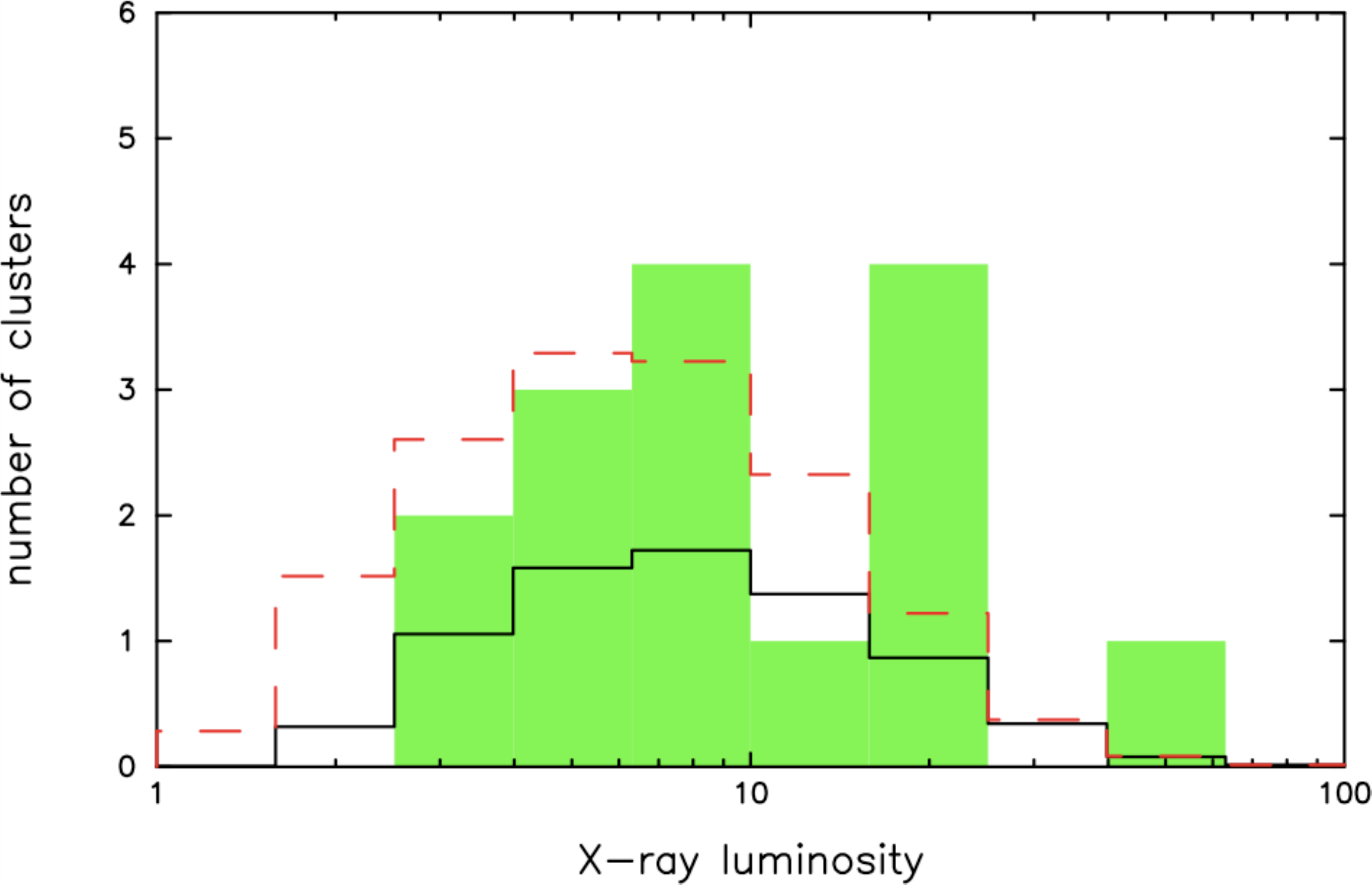}}
\subfloat[]{\label{fig:Yclusters}
\includegraphics[width=8cm]{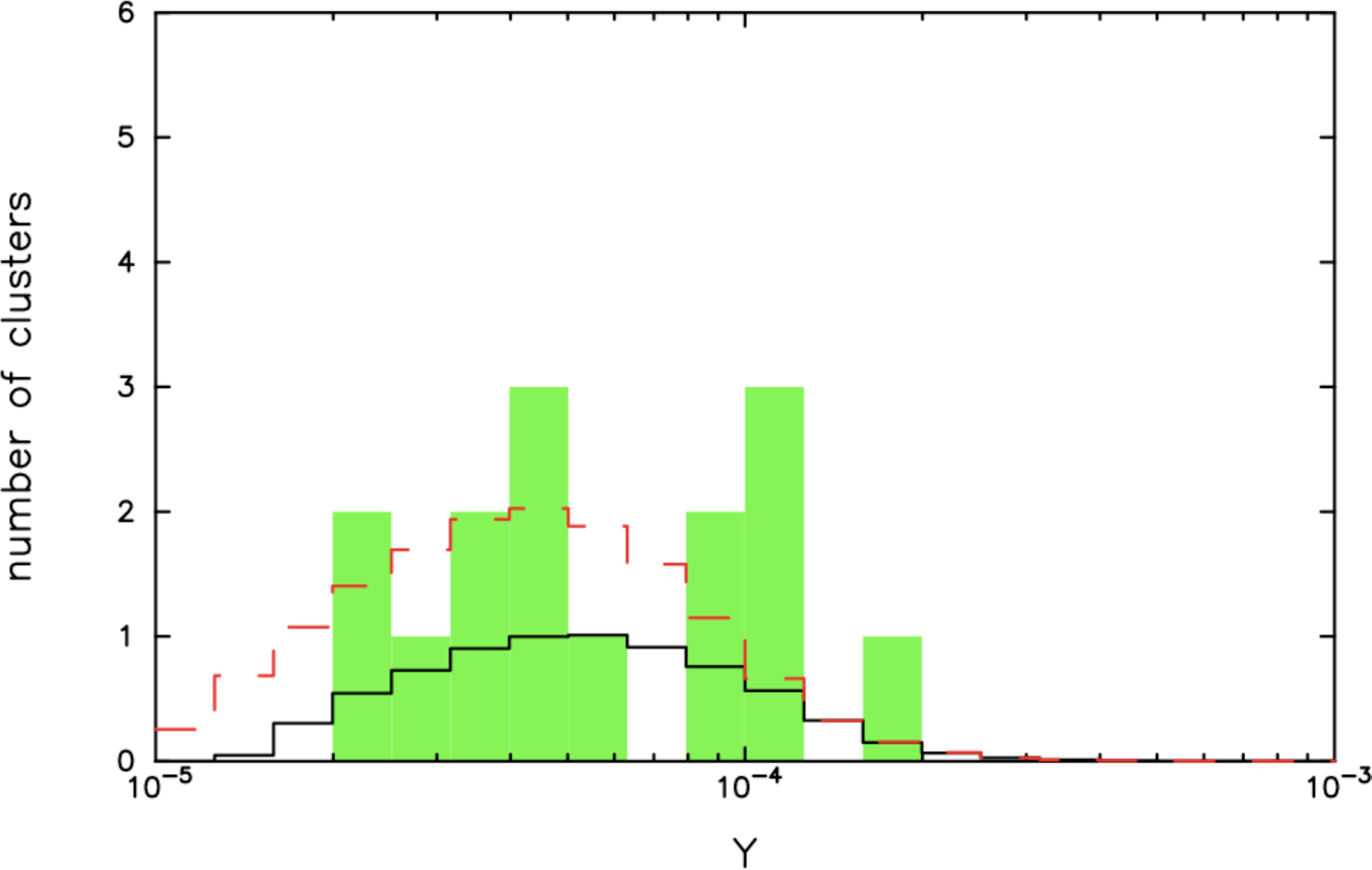}}
\caption{Mean distribution of the properties of the galaxy clusters 
that appear in the 20,000 randomly generated mock catalogues 
of the expected overlap between the XCS and {\it Planck} Cluster 
Survey (nominal mission - black/full, extended mission - red/long-dashed) 
and properties of the galaxy clusters presented in Table~1 (green, filled): 
(a) redshift; (b) X-ray temperature, $T_X$ (in units of keV); 
(c) X-ray bolometric luminosity,  $L_{\rm bol}$ (in units $10^{44}\;{\rm erg\,s^{-1}}$); 
(d) integrated comptonization parameter, $Y$ (in units $h^{-2}\;{\rm Mpc^{2}}$). 
Both $L_{\rm bol}$ and $Y$ are defined within $r_{200}$.}
\end{figure*}

However, if we assume the XCS will keep finding clusters suitable 
for {\it Planck} detection at the same rate as up until now, then 
the number of clusters in Table~1, which correspond to an effective 
survey sky area of $410$ square degrees, suggest the expected 
overlap between the final XCS catalogue and {\it Planck} 
Cluster Survey should be 16/22 clusters (nominal/extended cases). 
But such high numbers were only recovered in 68/1021 
(nominal/extended cases) out of the 20,000 random mock 
catalogues that were generated of the expected overlap between 
the XCS and {\it Planck} Cluster Survey. The mismatch 
between the theoretical and the empirical estimates cannot be 
solved as easily as one could imagine, given the uncertainty 
associated with all the priors that influence the outcome of the 
theoretical predictions. This is because modifying most of those 
priors leads to changes in the theoretical and the empirical 
estimates that go in the same way. For example, assuming 
better sensitivity for {\it Planck} would lead to an increase in  
the theoretical expectation for the overlap between the XCS 
and {\it Planck} Cluster Survey as well as in the number of 
clusters in Table~1. The same would happen if the assumed 
normalizations of the scaling relations between cluster 
observables and cluster mass were increased. However, 
increasing the value assumed for $\sigma_{8}$, or assuming 
the XCS to be more sensitive, would lead to an increase in 
the theoretical expectation for the overlap between the XCS 
and {\it Planck} Cluster Survey without changing much the 
number of clusters in Table~1. 

Alternatively, the X-ray luminosity of some of the galaxy clusters 
in Table~1 could have been increased by the presence of cool 
cores or mergers, whose effects may not have been properly 
modeled within the Millennium Gas Simulations, as well as 
due to AGN contamination. This would lead to an overestimation 
of $Y$, and thus also of the probability of {\it Planck} detection, 
because $Y$ is estimated from the pre-heating MGS scaling 
relations, given the observed value for the X-ray luminosity 
(as well as redshift and temperature) for each cluster. 

Our simulations suggest that {\it Planck} will detect about 630 
galaxy clusters, with X-ray temperatures in excess of 2 keV and in the
redshift interval $0.1 < z < 1.0$, in the nominal mission, increasing to 
approximately 1300 for the extended mission. This assumes the 
41,253 square degrees of sky  coverage adopted by \cite{Melin}. 
Although our result is lower than what others have obtained for the 
{\it Planck} nominal mission \citep{Melin,Bart,Leach,Chamb}, due to 
the different tools and assumptions made, the result for the scaling 
with mission duration that we obtained should be robust.

The results so far presented suggest that the information contained 
in just the XCS and {\it Planck} data will not allow a comprehensive 
characterization of the cross-correlations between the most 
important cluster X-ray and SZ properties, as a function of redshift. 
This will only be achieved if more data is gathered, through extensive 
follow-up of both surveys: they complement each other in terms of area 
and depth, while suffering from different selection biases. The XCS is, 
on average, deeper than the {\it Planck} Cluster Survey, and thus 
the XCS will contain almost all clusters detected by {\it Planck} in the 
sky area that is common to both surveys. Our simulations suggest that, 
at $z~\sim0.1$, almost 90 per cent of the galaxy clusters, with 
X-ray temperatures in excess of 2 keV, expected to be detected 
by {\it Planck} (in both the nominal and extended missions) will also 
be detected by the XCS (with more than 50 photon counts), slowly 
decreasing to slightly more than 70 per cent at $z\sim1$ (due to the 
existence of  observations with low exposure times in the {\it XMM} 
archive). This means that the XCS will not only be able to help 
better characterise the sensitivity of the {\it Planck} Cluster Survey, 
which is essential in order to recover the scaling relations pertaining 
to the underlying galaxy cluster population, but will also enable 
{\it Planck} to recover some information on many of the galaxy clusters 
that are just below its detection threshold in the sky area that is 
common to both surveys. 

\section{Conclusions}
\label{Conclusions}

We have characterised the galaxy clusters, with X-ray temperatures 
in excess of 2 keV and in the redshift interval $0.1 < z < 1.0$, that are 
expected to be simultaneously detected by the XCS (with more than 
250 X-ray photon counts) and {\it Planck} Cluster Survey. This overlap 
amounts to about 7/15 clusters for the nominal/extended {\it Planck} 
missions, taking the flat $\Lambda$CDM cosmology most compatible 
with WMAP-7 data and with the covariance of the cluster X-ray luminosity, 
temperature, and integrated comptonization parameter, as a function 
of cluster mass and redshift, determined by the pre-heating 
Millennium Gas Simulations. Under these 
assumptions, we have identified 11/15 clusters of galaxies in the first 
data release of the XCS catalogue, with a minimum of 250 X-ray photon 
counts, X-ray temperatures in excess of 2 keV and in the redshift interval 
$0.1 < z < 1.0$, that have more than a 50 per cent chance of being 
detected by {\it Planck} at the end of its nominal/extended missions. 
Both results clearly show that {\it Planck} will be able to detect only 
the closest, most luminous and hot clusters in the XCS. 

The galaxy clusters that appear in both the XCS and {\it Planck} 
Cluster Survey will provide valuable insights into the cross-correlations  
of X-ray and SZ properties, but their small number will not allow for 
the comprehensive characterization of such correlations. However, 
being on average the deeper of the two surveys, the XCS will help 
determine better the selection function of the {\it Planck} Cluster Survey. 
In the sky area that is common to both surveys, the XCS will also make 
possible the recovery of some information on many of the galaxy clusters 
that are just below the {\it Planck} detection threshold \citep{Planck3}, 
as well as significantly improve the $Y$ estimates for all clusters detected by 
{\it Planck} by aiding in the determination of cluster size \citep{Planck2}.
 
\section*{Acknowledgments}

This work was made possible by the ESA \emph{XMM-Newton} mission, 
and we thank everyone who was involved in making that mission such 
a success. PTPV and AdS acknowledge financial support from project 
PTDC/CTE-AST/64711/2006, funded by Funda\c{c}\~{a}o para a Ci\^{e}ncia 
e a Tecnologia. AdS was supported by a Ci\^{e}ncia 2007 contract, funded 
by FCT/MEC (Portugal) and POPH/FSE (EU). EPRGR was financially 
supported by a grant from Funda\c{c}\~ao para a Ci\^encia e a Tecnologia 
(POPH-QREN-SFRH/BD/45613/2008). AKR, ARL, ELD, MHo, MS, NM, 
were supported by the Science and Technology Facilities Council 
(STFC) [grants number ST/F002858/1 and ST/I000976/1].
MHo acknowledges financial support from the Graduate Teaching Associate
programme at the School of Science and Technology at the University of
Sussex. NM acknowledges financial support from a PPARC/STFC
studentship. MHi acknowledges the support support from the 
Leverhulme Trust. MS acknowledges 
financial support from the Swedish Research Council (VR) through the Oskar 
Klein Centre. SAS notes that this work was performed under the auspices of the 
U.S. Department of Energy, National Nuclear Security Administration by the 
University of California, Lawrence Livermore National Laboratory under contract 
No. W-7405-Eng-48. We thank Jean-Baptiste Melin for discussions, and Gus Evrard 
and Rebecca Stanek for re-visiting their estimation of the MGS cluster scaling relations.



\bsp
\end{document}